\documentclass[aps,tightenlines,nofootinbib,preprint,
superscriptaddress,groupedaddress,epsfig]{revtex4}

\usepackage{graphicx}

\begin{document}

\title{Annihilation Rate of $2^{++}$ Charmonium and Bottomonium}
\vspace{2cm}

\author{Guo-Li Wang}
\email{gl_wang@hit.edu.cn} \affiliation{Department of Physics,
Harbin Institute of Technology, Harbin, 150001, China}

\baselineskip=20pt

\vspace{2cm}

\begin{abstract}

Two-photon annihilation rates of $2^+$ tensor charmonium and
bottomonium up to third radial excited states are estimated in the
relativistic Salpeter method. Full Salpeter equation for $2^+$
tensor state is solved with a well defined relativistic wave
function and we calculated the annihilation amplitude using the
Mandelstam formalism. Our estimates of the decay widths are:
$\Gamma(\chi_{c2} \rightarrow 2\gamma)=501$ eV, $\Gamma(\chi'_{c2}
\rightarrow 2\gamma)=534$ eV, $\Gamma(\chi_{b2} \rightarrow
2\gamma)=7.4$ eV and $\Gamma(\chi'_{b2} \rightarrow 2\gamma)=7.7$
eV. We also give total decay widths of the lowest states estimated
by the two-gluon decay rates, and the results are:
$\Gamma_{tot}(\chi_{c2})=2.64$ MeV,
$\Gamma_{tot}(\chi_{b2})=0.220$ MeV and
$\Gamma_{tot}(\chi'_{b2})=0.248$ MeV.

\end{abstract}


\maketitle

\section{Introduction}

Recently, the radiative annihilation physics of $\chi_{c0}$,
$\chi_{b0}$ ($0^{++}$), $\chi_{c2}$ and $\chi_{b2}$ ($2^{++}$)
become hot topics
\cite{Schuler,Ma,Huang,Lakhina,Crater,Mangano,Bodwin,Patrignani,Ebert,Munz,Gupta},
because the annihilation amplitudes are related to the behavior of
the wave functions, so the annihilation rates are helpful to
understand the formalism of inter-quark interactions, and can be a
sensitive test of the potential model \cite{Godfrey}.

In previous letter \cite{twophoton2}, two-photon and two-gluon
annihilation rates of $0^{++}$ scalar $c\bar c$ and $b \bar b$
states are computed in the relativistic Salpeter method, good
agreement of our predictions with other theoretical calculations
and available experimental data is found. In our calculation, we
found the relativistic corrections are large and can not be
ignored, and point out that all the calculations related to a $P$
wave state, one have to use a relativistic method, a
non-relativistic method will cause a large error, even for a heavy
state \cite{twophoton2,decayconstant}. In a non-relativistic
calculation, the corresponding decay width is related to the
derivative of the non-relativistic P-wave function at the origin,
but in a full relativistic calculation, the relativistic
corrections include not only relativistic kinematics but also the
relativistic inter-quark dynamics, the decay width is related to
the full behavior of P-wave function which can be seen in this
letter or in Ref. \cite{twophoton}.

 In this letter, we give relativistic calculation of $2^{++}$
 tensor $c\bar c$ and $b \bar b$ states decaying into two
photons using the instantaneous Bethe-Salpeter method \cite{BS},
which is a full relativistic Salpeter method \cite{salp}. The case
of tensor $2^{++}$ state is special, not like other P-wave states,
there are no decay constants in this state, and because there is
the $P-F$ mixing problem, it make the physics much complicated.

The reason of the exist of $P-F$ mixing is that $P$ wave and $F$
wave state have the same parity and charge conjugate parity, they
are all $J^{PC}=2^{++}$, one can not distinguish them by the
quantum number. We have found a basic method to deal with this
problem \cite{vector}, we begin from the quantum field theory,
analyze the parity and charge conjugation of bound state, and give
a formula for the wave function that is in a relativistic form
with definite parity and charge conjugation symmetry, then we
solve the full Salpeter equation, and obtain the corresponding
state, and there are automatically the mixing between $P$ wave and
$F$ wave.

The letter is organize as following, In Sec.II, we give the
annihilation amplitude in Mandelstam formalism and the wave
function of the $2^+$ tensor state with a well defined
relativistic form. The two-photon decay width and full width of
heavy $2^{++}$ quarkonium are formulated in Sec.II, we show the
numerical results and give discussions in the Sec.III.

\section{Theoretical Details}

\begin{figure}
\begin{picture}(250,130)(200,400)
\put(0,0){\includegraphics{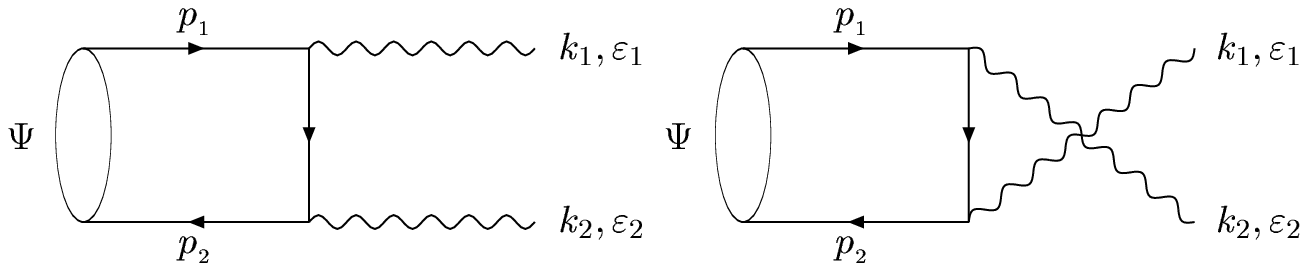}}
\end{picture}
\caption{Two-photon annihilation diagrams of the quarkonium. }
\end{figure}

According to the Mandelstam \cite{mandelstam} formalism, the
relativistic transition amplitude of a quarkonium decaying into
two photons (see figure 1) can be written as:
\begin{eqnarray}
T_{2\gamma} & = & i\sqrt{3}\; (iee_q)^2 \!
 \int \!\! \frac{d^4q}{(2\pi)^4}
              \; \mbox{tr}\; \Bigg\{ \,
    \chi(q) \bigg[\varepsilon\!\!\! /_2\, S(p_1-k_1)\,
    \varepsilon\!\!\! /_1 +
     \varepsilon\!\!\! /_1\, S(p_1-k_2)\,
    \varepsilon\!\!\! /_2 \bigg]
        \Bigg\},
\label{eq1}
\end{eqnarray}
where $k_1$, $k_2$; $\varepsilon_1$, $\varepsilon_2$ are the
momenta and polarization vectors of photon 1 and photon 2;
$e_q=\frac{2}{3}$ for charm quark and $e_q=\frac{1}{3}$ for bottom
quark; $p_1$ and $p_2$ are the momentum of constitute quark 1 and
antiquark 2; $\chi(q)$ is the Bethe-Salpeter wave function of the
corresponding meson with the total momentum $P$ and relative
momentum $q$, related by
$$p_{_1}={\alpha}_{1}P+q, \;\; {\alpha}_{1}\equiv\frac{m_{1}}{m_{1}+m_{2}}~,$$
$$p_{2}={\alpha}_{2}P-q, \;\; {\alpha}_{2}\equiv\frac{m_{2}}{m_{1}+m_{2}}~,$$
where $m_1=m_2$ is the constitute quark mass of charm or bottom.

Since $p_{10}+p_{20}=M$, the approximation
$p_{10}=p_{20}=\frac{M}{2}$ is a good choice for the equal mass
system \cite{barbieri,keung,Huang}. Having this approximation, we
can perform the integration over $q_0$ to reduce the expression,
with the notation of Salpeter wave function $\Psi(\vec
q)=\int\frac{dq_0}{2\pi}\chi(q)$, to
\begin{eqnarray}
T_{2\gamma}  =  \sqrt{3}\; (ee_q)^2 \!
 \int \!\! \frac{d\vec{q}}{(2\pi)^3}
              \; \mbox{tr}\; \Bigg\{ \,
    \Psi(\vec q) \bigg[\varepsilon\!\!\! /_2\, \frac{1}{\not\! p_1-\not\! k_1-m_1}\,
    \varepsilon\!\!\! /_1 +
     \varepsilon\!\!\! /_1\, \frac{1}{\not\! p_1-\not\! k_2-m_1}\,
    \varepsilon\!\!\! /_2 \bigg]
        \Bigg\}.
\label{eq2}
\end{eqnarray}

The general form for the relativistic wave function of tensor
$J^{P}=2^{+}$ state (or $J^{PC}=2^{++}$ for equal mass system) can
be written as $16$ terms constructed by momentum $P$, $q$ and
Dirac matrix $\gamma$, because of the approximation of
instantaneous, $8$ terms with $P\cdot q$ become zero, the
relativistic Salpeter wave function $\Psi(\vec q)$ for $2^{+}$
state can be written as:

$$\Psi_{2^{+}}(\vec{q})=
{\varepsilon}_{\mu\nu}{q_{\perp}^{\nu}}
\left\{{q_{\perp}^{\mu}}\left[f_1(\vec{q})+\frac{\not\!P}{M}f_2(\vec{q})+
\frac{{\not\!q}_{\perp}}{M}f_3(\vec{q})+\frac{{\not\!P}
{\not\!q}_{\perp}}{M^2} f_4(\vec{q})\right]\right.$$
\begin{equation}\left.+
{\gamma^{\mu}}\left[Mf_5(\vec{q})+ {\not\!P}f_6(\vec{q})+
{\not\!q}_{\perp} f_7(\vec{q})\right]+\frac{i}{M}
f_8(\vec{q})\epsilon^{\mu\alpha\beta\gamma}
P_{\alpha}q_{\perp\beta}\gamma_{\gamma}\gamma_{5}\right\},\label{eq3}
\end{equation}
where the ${\varepsilon}_{\mu\nu}$ is the polarization tensor of
the $2^{+}$ state, ${q}_{\perp}=(0,\vec{q})$. But these 8 terms
wave functions $f_i$ are not independent, there are the further
constraint from Salpeter equation \cite{salp}: $
\Psi^{+-}_{2^{+}}(\vec{q})=\Psi^{-+}_{2^{+}}(\vec{q})=0\;, $ which
give the constraints on the components of the wave function, so we
get the relations
$$f_1(\vec{q})=\frac{\left[q_{\perp}^2 f_3(\vec{q})+M^2f_5(\vec{q})
\right](\omega_1+\omega_2)-M^2f_5(\vec{q})(\omega_1-\omega_2)}
{M(m_1\omega_2+m_2\omega_1)},~$$
$$f_2(\vec{q})=\frac{\left[q_{\perp}^2 f_4(\vec{q})-M^2f_6(\vec{q})\right]
(\omega_1-\omega_2)}
{M(m_1\omega_2+m_2\omega_1)},~~$$\begin{equation}~~f_7(\vec{q})=\frac{f_5(\vec{q})M(\omega_1-\omega_2)}
{m_1\omega_2+m_2\omega_1},~~~~f_8(\vec{q})=\frac{f_6(\vec{q})M(\omega_1+\omega_2)}
{m_1\omega_2+m_2\omega_1}.\end{equation} Only four independent
wave functions $f_3(\vec{q})$, $f_4(\vec{q})$, $f_5(\vec{q})$ and
$f_6(\vec{q})$ been left, one can check in Eq.(\ref{eq3}), all the
terms except the two terms with $f_2$ and $f_7$ are charge
conjugate parity positive, but $f_2$ and $f_7$ terms have negative
charge conjugate parity, after we use the constraint relations,
for equal mass system, the terms with $f_2$ and $f_7$ become zero,
then the whole wave function have positive charge conjugate
parity, that is $2^{++}$ state. These wave functions and the bound
state mass $M$ can be obtained by solving the full Salpeter
equation with the constituent quark mass as input, and we will not
show the details of how to solve it, we only show our result in
next section.

These four independent wave functions fulfil the normalization
condition:
$$\int \frac{d{\vec q}}{(2\pi)^3}\frac{16~\omega_1\omega_2~{\vec
q}^2}{15(m_1\omega_2+m_2\omega_1)}\left\{
f_5~f_6~M^2\left[5+\frac{(m_1+m_2)(m_2\omega_1-m_1\omega_2)}
{\omega_1\omega_2(\omega_1+\omega_2)}\right]\right.
$$\begin{equation}\left.
+f_4~f_5~{\vec
q}^2\left[2+\frac{(m_1+m_2)(m_2\omega_1-m_1\omega_2)}
{\omega_1\omega_2(\omega_1+\omega_2)}\right] -2~{\vec
q}^2f_3\left(f_4\frac{{\vec q}^2}{M^2}+f_6\right) \right\}=2M.
 \end{equation}

With the full Salpeter wave function, the two photon decay
amplitude can be written as:
$$T_{2\gamma}=4~\sqrt{3}~e^2~e^2_q ~{\varepsilon}_{\mu\nu} \int \frac{d{\vec
q}}{(2\pi)^3}\left\{ \frac{1}{x_1-|{\vec
q}|~M~\cos\theta}\left[~f_5 ~M~\left(
{k_{1}^{\mu}}~{q_{\perp}^{\nu}}~
{\varepsilon}_1\cdot{\varepsilon}_2 \right.\right.\right.$$

$$\left.\left.  +
{\varepsilon}_1^{\mu}~{q_{\perp}^{\nu}}~{\varepsilon}_2\cdot
{q_{\perp}}+
{\varepsilon}_2^{\mu}~{q_{\perp}^{\nu}}~{\varepsilon}_1\cdot
{q_{\perp}}\right)+\frac{{q_{\perp}^{\mu}}~{q_{\perp}^{\nu}}~f_3}{M}
\left({\varepsilon}_1\cdot{\varepsilon}_2~q_{\perp}\cdot
k_1+2~{\varepsilon}_1\cdot {q_{\perp}}~{\varepsilon}_2\cdot
{q_{\perp}}\right)\right]$$

$$+ \frac{1}{x_1+|{\vec q}|~M~\cos\theta}\left[~f_5
~M~\left( {k_{2}^{\mu}}~{q_{\perp}^{\nu}}~
{\varepsilon}_1\cdot{\varepsilon}_2+
{\varepsilon}_1^{\mu}~{q_{\perp}^{\nu}}~{\varepsilon}_2\cdot
{q_{\perp}} +
{\varepsilon}_2^{\mu}~{q_{\perp}^{\nu}}~{\varepsilon}_1\cdot
{q_{\perp}}\right) \right.$$

\begin{equation}\left.\left.+\frac{{q_{\perp}^{\mu}}~{q_{\perp}^{\nu}}~f_3}{M}
\left({\varepsilon}_1\cdot{\varepsilon}_2~q_{\perp}\cdot
k_2+2~{\varepsilon}_1\cdot {q_{\perp}}~{\varepsilon}_2\cdot
{q_{\perp}}\right)\right]\right\}\end{equation}

where $x_1={\frac{M^2}{4}+{\vec q}^2+m^2_1}$, and $\theta$ is the
angle between the momentum $\vec q$ and $\vec k_1$. Finally the
decay width with first order QCD correction \cite{Kwong} can be
written as:
\begin{equation}\Gamma_{2\gamma}=\frac{1}{2!\cdot 5\cdot 16~\pi M}~
\sum|T_{2\gamma}|^2\cdot\left(1-\frac{16\alpha_s}{3\pi}\right).\end{equation}

Until now, only the total decay width of $\chi_{c2}(1P)$ is
available, we can estimate the full decay width of OZI-forbidden
states using the two gluon decay width, and the two gluon decay
width of quarkonium can be easily obtained from the two photon
decay width with a simple replacement. For the charmnium states,
only the ground state $\chi_{c2}(1P)$ is OZI-forbidden, we have:
\begin{equation}\Gamma_{tot}(\chi_{c2})\cong\Gamma_{2g}(\chi_{c2})=
\Gamma_{2\gamma}(\chi_{c2})~\frac{{9~\alpha^2_s(m_c)}}
{8~\alpha^2}~\frac{1-\frac{2.2~\alpha_s(m_c)}{\pi}}
{1-\frac{16~\alpha_s(m_c)}{3~\pi}}.\end{equation}

For $2^{++}$ bottomnium states, according to our estimate of mass
spectra, there are two states which are below the threshold of
$B\bar B$, and we can predict their full decay widths using their
two gluon decay widths. For $\chi_{b2}(1P)$, we have:
\begin{equation}\Gamma_{tot}(\chi_{b2})\cong\Gamma_{2g}(\chi_{b2})=
\Gamma_{2\gamma}(\chi_{b2})~\frac{{18~\alpha^2_s(m_b)}}
{\alpha^2}~\frac{1-\frac{0.1~\alpha_s(m_b)}{\pi}}
{1-\frac{16~\alpha_s(m_b)}{3~\pi}},\end{equation} and for the
first radial excited state $\chi'_{b2}(2P)$, we have:
\begin{equation}\Gamma_{tot}(\chi'_{b2})\cong\Gamma_{2g}(\chi'_{b2})=
\Gamma_{2\gamma}(\chi'_{b2})~\frac{{18~\alpha^2_s(m_b)}}
{\alpha^2}~\frac{1+\frac{\alpha_s(m_b)}{\pi}}
{1-\frac{16~\alpha_s(m_b)}{3~\pi}},\end{equation} where
$\alpha=\frac{e^2}{4\pi}$, and the QCD corrections are summarized
in Ref.\cite{Kwong}.

\section{Numerical Results and Discussions}

We will not show the details of Solving the full Salpeter
equation, only give the final results, interested readers can find
the the detail technique in Ref.~\cite{cskimwang}.

When solving the full Salpeter equation, we choose a
phenomenological Cornell potential,
$$V(\vec q)=V_s(\vec q)
+\gamma_{_0}\otimes\gamma^0 V_v(\vec q)~,$$
$$V_s(\vec q)=-(\frac{\lambda}{\alpha}+V_0)
\delta^3(\vec q)+\frac{\lambda}{\pi^2} \frac{1}{{(\vec
q}^2+{\alpha}^2)^2}~,$$
\begin{equation}
V_v(\vec q)=-\frac{2}{3{\pi}^2}\frac{\alpha_s( \vec q)}{{(\vec
q}^2+{\alpha}^2)}~,
\end{equation}
and the coupling constant $\alpha_s(\vec q)$ is running:
$$\alpha_s(\vec q)=\frac{12\pi}{25}\frac{1}
{\log (a+\frac{{\vec q}^2}{\Lambda^{2}_{QCD}})}~.$$ Here the
constants $\lambda$, $\alpha$, $a$, $V_0$ and $\Lambda_{QCD}$ are
the parameters that characterize the potential. We found the
following best-fit values of input parameters which were obtained
by fitting the mass spectra for $2^{++}$ $\chi_{c2}$:
$a=e=2.7183$, $\alpha=0.06$ GeV, $V_0=-0.401$ GeV, $\lambda=0.2$
GeV$^2$, $\Lambda_{QCD}=0.26$ GeV and
 $m_c=1.7553$ GeV.
With this parameter set, we solve the full Salpeter equation and
obtain the mass spectra shown in Table I. To give the numerical
value, we need to fix the value of the renormalization scale $\mu$
in $\alpha_s(\mu)$. In the case of charmonium, we choose the charm
quark mass $m_c$ as the energy scale and obtain the coupling
constant $\alpha_s(m_c)=0.36$ \cite{cskimwang}.

Since the wave function Eq.(\ref{eq3}) is general wave function
for $2^+$ state, and either $^3P_2$ and $^3F_2$ can be $2^+$
state, so the obtained states of $\chi^{'}_{c2}(1F)$ and
$\chi^{''}_{c2}(2F)$ are not pure $F$ wave, they are $P-F$ mixing
state, but a $F$ wave dominate state, and $\chi^{'}_{b2}(1F)$,
$\chi^{''}_{b2}(2F)$ in Table II are also $P-F$ mixing, but $F$
wave dominate state, we will discuss the detail of mixing in other
paper\cite{new}.

Our prediction of the mass for $\chi'_{c2}(2P)$ is $3967.0$ MeV,
which is a little larger than the first observation by BELLE
collaboration, their data is $3931$ MeV. And our prediction of the
first $F$ wave dominate state, we list as $\chi^{'}_{c2}(1F)$,
whose mass is $4040.5$ MeV, and the second one,
$\chi^{''}_{c2}(2F)$, whose mass is $4314.3$ MeV.

With the obtained wave function and Eq.(\ref{eq3}), we calculate
the two-photon decay width of $c\bar c$ $2^{++}$ states, the
results are also shown in Table I. Not similar to the $S$ wave
case, where the two photon decay width of ground state
$\Gamma(\eta_c \rightarrow 2\gamma)$ is much larger then the first
radial excited state $\Gamma(\eta'_c \rightarrow 2\gamma)$, almost
twice\cite{twophoton}, the ground state decay width is a little
smaller than the first radial excited state, and from the table we
obtain the conclusion that the decay widths with successive radial
excitation fall very slowly. The decay width of the $F$ wave
dominate state, $\Gamma(\chi^{'}_{c2}(1F)\rightarrow
2\gamma)=92~{\rm eV}$ and $\Gamma(\chi^{''}_{c2}(2F)\rightarrow
2\gamma)=54~{\rm eV}$ are much smaller than the $P$ wave dominate
state.

For the case of $b \bar b$, our best fitting parameters are
$V_0=-0.459$ GeV, $\Lambda_{\rm QCD}=0.20$ and $m_b=5.13$ GeV,
other parameters are same as in the case of $c \bar c$. With this
set of parameters, the coupling constant at scale of bottom quark
mass is $\alpha_s(m_b)=0.232$. The corresponding mass spectra,
two-photon decay widths are shown in Table II. Our mass prediction
of $\chi'_{b2}(2P)$ is $10283$ MeV, a little higher than the data.
And the two photon decay widths are much smaller than the case of
charmonium, their value is only about several eV.

\begin{table*}[hbt]
\setlength{\tabcolsep}{0.5cm} \caption{\small Two-photon decay
width and total width of P-wave $2^{++}$ charmonium states, where
the data of $\chi_{c2}(1P)$ is come from PDG\cite{PDG}, and data
of $\chi^{'}_{c2}(2P)$ is come from Ref.\cite{belle}.}
\label{tab1}
\begin{tabular*}{\textwidth}{@{}c@{\extracolsep{\fill}}cccc}
 \hline \hline
{\phantom{\Large{l}}}\raisebox{+.2cm}{\phantom{\Large{j}}}
&Ex~Mass~(MeV)&Th~Mass~(MeV)&$\Gamma_{2\gamma}$~~(eV )\\
\hline\hline
{\phantom{\Large{l}}}\raisebox{+.2cm}{\phantom{\Large{j}}}
$\chi_{c2}(1P)$~&3556.20&~3556.6~&~501 \\
{\phantom{\Large{l}}}\raisebox{+.2cm}{\phantom{\Large{j}}}
$\chi^{'}_{c2}(2P)$~&3931&~3967.0~&~534 \\
{\phantom{\Large{l}}}\raisebox{+.2cm}{\phantom{\Large{j}}}
$\chi^{'}_{c2}(1F)$~&&~4040.5~&~92.4 \\
{\phantom{\Large{l}}}\raisebox{+.2cm}{\phantom{\Large{j}}}
$\chi^{''}_{c2}(3P)$~&&~4264.6~&~509 \\
{\phantom{\Large{l}}}\raisebox{+.2cm}{\phantom{\Large{j}}}
$\chi^{''}_{c2}(2F)$~&&~4314.3~&~54.1 \\
{\phantom{\Large{l}}}\raisebox{+.2cm}{\phantom{\Large{j}}}
$\chi^{'''}_{c2}(4P)$~&&~4498.7~&~475
 \\\hline\hline
\end{tabular*}
\end{table*}

\begin{table*}[hbt]
\setlength{\tabcolsep}{0.5cm} \caption{\small Two-photon decay
width and total width of P-wave $2^{++}$ bottomonium states.}
\label{tab2}
\begin{tabular*}{\textwidth}{@{}c@{\extracolsep{\fill}}cccc}
 \hline \hline
{\phantom{\Large{l}}}\raisebox{+.2cm}{\phantom{\Large{j}}}
&Ex~Mass~(MeV)&Th~Mass~(MeV)&$\Gamma_{2\gamma}$~~(eV) \\
\hline\hline
{\phantom{\Large{l}}}\raisebox{+.2cm}{\phantom{\Large{j}}}
$\chi_{b2}(1P)$&9912.21&~9912.2~&~7.43 \\
{\phantom{\Large{l}}}\raisebox{+.2cm}{\phantom{\Large{j}}}
$\chi^{'}_{b2}(2P)$~&10268.65&~10283~&~7.69 \\
{\phantom{\Large{l}}}\raisebox{+.2cm}{\phantom{\Large{j}}}
$\chi^{'}_{b2}(1F)$~&&~10364~&~1.76 \\
{\phantom{\Large{l}}}\raisebox{+.2cm}{\phantom{\Large{j}}}
$\chi^{''}_{b2}(3P)$~&&~10561~&~7.19 \\
{\phantom{\Large{l}}}\raisebox{+.2cm}{\phantom{\Large{j}}}
$\chi^{''}_{b2}(2F)$~&&~10616~&~1.43 \\
{\phantom{\Large{l}}}\raisebox{+.2cm}{\phantom{\Large{j}}}
$\chi^{'''}_{b2}(4P)$~&&~10786~&~6.60
 \\\hline\hline
\end{tabular*}
\end{table*}

\begingroup
\squeezetable
\begin{table*}[hbt]
\setlength{\tabcolsep}{0.5cm} \caption{\small Recent theoretical
and experimental results of two-photon decay width and total
width.} \label{tab3}
\begin{tabular*}{\textwidth}{@{}c@{\extracolsep{\fill}}ccccccc}
 \hline \hline
{\phantom{\Large{l}}}\raisebox{+.2cm}{\phantom{\Large{j}}}
&$\Gamma^{\chi_{c2}}_{2\gamma}$~(keV)
~&$\Gamma^{\chi_{c2}}_{tot}$~(MeV)~&
$\Gamma^{\chi^{'}_{c2}}_{2\gamma}$~(keV)~&
$\Gamma^{\chi_{b2}}_{2\gamma}$~(eV)~
&$\Gamma^{\chi_{b2}}_{tot}$~(MeV)~&
$\Gamma^{\chi^{'}_{b2}}_{2\gamma}$~(eV)~\\\hline\hline
{\phantom{\Large{l}}}\raisebox{+.2cm}{\phantom{\Large{j}}}
Ours~&~0.50 &2.64&0.53&7.4&0.22&7.7\\
 {\phantom{\Large{l}}}\raisebox{+.2cm}{\phantom{\Large{j}}}
Gupta \cite{Gupta}  ~&~0.57&1.20&&8&0.22&\\
{\phantom{\Large{l}}}\raisebox{+.2cm}{\phantom{\Large{j}}}
Ebert \cite{Ebert} ~&~0.50 &&0.52&8&&6&\\
{\phantom{\Large{l}}}\raisebox{+.2cm}{\phantom{\Large{j}}}
 M\"unz \cite{Munz} ~&~0.44$\pm$0.14 &&0.48$\pm$0.16&5.6$\pm$0.6&&6.8$\pm$1.0\\
 {\phantom{\Large{l}}}\raisebox{+.2cm}{\phantom{\Large{j}}}
 CLEO \cite{CLEO01} ~&~0.53$\pm$0.15$\pm$0.06$\pm$0.22&&&&&\\
 {\phantom{\Large{l}}}\raisebox{+.2cm}{\phantom{\Large{j}}}
 CLEO \cite{CLEO05} ~&~0.56$\pm$0.06$\pm$0.05$\pm$0.04&&&&&\\
 {\phantom{\Large{l}}}\raisebox{+.2cm}{\phantom{\Large{j}}}
  CLEO \cite{CLEO08} ~&~0.60$\pm$0.06$\pm$0.06&&&&&\\
 {\phantom{\Large{l}}}\raisebox{+.2cm}{\phantom{\Large{j}}}
  PDG \cite{PDG} ~&~0.493 &2.03$\pm$0.12&&&&\\

 \hline\hline
\end{tabular*}
\end{table*}
\endgroup

Our predictions of the total decay width for the ground state:
$$\Gamma_{tot}(\chi_{c2})\cong\Gamma_{2g}(\chi_{c2})=2.64~{\rm MeV},$$ is
little larger then the PDG data
$\Gamma_{tot}(\chi_{c2})=2.03\pm0.12$ MeV. For $2^{++}$ bottomnium
states, we have no data in hand, our theoretical predictions are:
$$\Gamma_{tot}(\chi_{b2})\cong\Gamma_{2g}(\chi_{b2})=0.220~{\rm MeV}$$ and
$$\Gamma_{tot}(\chi'_{b2})\cong\Gamma_{2g}(\chi'_{b2})=0.248~{\rm MeV}.$$

We compare our predictions with recent other theoretical
relativistic calculations and experimental results in Table III.
Except the total decay width of $\Gamma_{tot}(\chi_{c2})$, all the
values of
 listed in the table consist with each
other.

In summary, by solving the relativistic full Salpeter equation
with a well defined form of wave function, we estimate two-photon
decay rates: $\Gamma(\chi_{c2} \rightarrow 2\gamma)=501$ eV,
$\Gamma(\chi'_{c2} \rightarrow 2\gamma)=534$ eV, $\Gamma(\chi_{b2}
\rightarrow 2\gamma)=7.4$ eV and $\Gamma(\chi'_{b2} \rightarrow
2\gamma)=7.7$ eV, and the total decay widths:
$\Gamma_{tot}(\chi_{c2})=2.64$ MeV,
$\Gamma_{tot}(\chi_{b2})=0.220$ MeV and
$\Gamma_{tot}(\chi'_{b2})=0.248$ MeV.\\

\newpage

\acknowledgements

This work was supported in part by the National Natural Science
Foundation of China (NSFC) under Grant No. 10675038 and No.
10875032, and in part by SRF for ROCS, SEM.
\\

\end{document}